\input{aipcheck}

\documentclass[
    ,final            
  ]
  {aipproc}

\layoutstyle{6x9}
\usepackage{epsfig,latexsym,amssymb,color,amsmath}
\usepackage{graphicx,graphics}
\usepackage[LGR,T1]{fontenc}
\usepackage{multirow, array}
\usepackage{soul,colortbl}
\DeclareGraphicsExtensions{.jpg,.pdf,.png}

\begin{document}

\title{A Double Ion Trap for Large Coulomb Crystals}


\author{Caroline Champenois}{
   address={Université d'Aix-Marseille, CNRS, PIIM, UMR7345\\ Centre de St Jérôme, Case C21, 13397 Marseille Cedex 20, France}
}

\author{Jofre Pedregosa-Gutierrez}{
 address={Université d'Aix-Marseille, CNRS, PIIM, UMR7345\\ Centre de St Jérôme, Case C21, 13397 Marseille Cedex 20, France}
}

\author{Mathieu Marciante}{
   address={Université d'Aix-Marseille, CNRS, PIIM, UMR7345\\ Centre de St Jérôme, Case C21, 13397 Marseille Cedex 20, France}
}

\author{Didier Guyomarc'h}{
  address={Université d'Aix-Marseille, CNRS, PIIM, UMR7345\\ Centre de St Jérôme, Case C21, 13397 Marseille Cedex 20, France}
}

\author{Marie Houssin}{
 address={Université d'Aix-Marseille, CNRS, PIIM, UMR7345\\ Centre de St Jérôme, Case C21, 13397 Marseille Cedex 20, France}
}

\author{Martina Knoop}{
 address={Université d'Aix-Marseille, CNRS, PIIM, UMR7345\\ Centre de St Jérôme, Case C21, 13397 Marseille Cedex 20, France}
}

\begin{abstract}
While the linear radiofrequency trap finds various applications in high-precision spectroscopy and quantum information, its higher-order cousin, the linear multipole trap, is almost exclusively employed in physical chemistry. Recently, first experiments have shown interesting features by laser-cooling  multipole-trapped ion clouds. Multipole traps show a flatter potential in their centre and therefore a modified density distribution compared to quadrupole traps. Micromotion is an important issue and will certainly influence the dynamics of crystallized ion structures. Our experiment tends to investigate possible crystallization processes in the multipole. In a more general way, we are interested in the study of the dynamics and thermodynamics of large ion clouds in traps of different geometry.
\end{abstract}

\maketitle


\section{Multipole traps}

Thirty years after Wolfgang Paul's introduction of the 3D radiofrequency (rf) ion cage, the linear version of the quadrupole trap has been proposed for use in high precision spectroscopy \cite{prestage89}. Further evolution to linear traps with an increased number of electrodes has been motivated by the profile of the local rf electric field which is no more linear with the distance to the trap axis but is ruled by a higher-order  law. As a consequence, for the same potential depth as in a quadrupole trap, multipole rf  traps offer a central trapping volume nearly free of rf electric field and therefore exempt of rf driven motion.  This is of major importance  in  physical chemistry \cite{gerlich92}, where the kinetic energy of the collision must be controlled and ideally kept very low for cold collisional reaction studies. The cryogenic 22-pole trap is widely used to study ion-molecule reactions and complex molecular spectroscopy, a recent review of this trap and its applications can be found in \cite{wester09}.

For this type of trap, only one application in frequency metrology is reported which takes advantage of the nearly field free zone offered by linear multipole traps to reduce the rf induced Doppler effect \cite{prestage07}.  In this device,  the multipole trap is combined in line with a quadrupole trap, using a shuttling protocol between both parts in order to combine the advantages of both confinement zones.
Recently, first experimental results have been reported on laser-cooling in an octupole and in a hexapole trap \cite{okada07,okada09}, limited however by experimental constraints. Numerical simulations \cite{okada07} have demonstrated that laser-cooled ions form stable structures, called Coulomb crystals,  in these traps. Due to the increase of the particle density  with the distance from the axis expected for cold samples in multipole traps \cite{champenois09}, Coulomb crystals show a hollow core structure \cite{okada09},  very different from what is observed in quadrupole rf traps where the ion density of cold sample is uniform \cite{hornekaer02}. The stability of hollow core Coulomb crystal structures remains to be studied, under the competing influences of rf driven motion (micromotion) and laser-cooling.

We have designed and realized an experimental device which is dedicated to the study of stable ion structures when trapped in a  multipole geometry.
In the following we start by giving a brief introduction to the theory that rules the dynamics of ions stored in a linear multipole trap. We  then present some of the studies that have been carried out numerically in these systems. The experimental device and first results are described in the last part of this paper.

\section{Theory of the multipole trap}
We assume a linear multipole trap made of $2k$ rods where  a radio-frequency voltage of peak-to-peak amplitude $V_0$ and frequency $\Omega/2\pi$ is applied to the rods of the trap, with exact phase opposition between adjacent rods. The potential along the $z$-axis is created by applying a DC-voltage $V_{end}$ onto the end-cap electrodes. For reasons of simplicity, we assume that no DC-voltages are applied on the rf rods.
The total effective time-averaged potential (or pseudopotential) in this device is then
\begin{equation}
V^*(\mathbf{r})=\frac{q^2V_0^2}{32 {\mathcal E}_k}\left(\frac{r}{r_0}\right)^{2k-2} + \ \frac{q\kappa V_{end}}{2z_0^2}(2z^2-r^2),
\end{equation}
where ${\mathcal E}_k=m\Omega^2 r_0^2/(2 k^2)$ is a characteristic energy, $\kappa$ is a loss factor depending on the geometry of the end electrodes relative to the rods. $\kappa$ includes all screening and geometric effects that  explain the difference between the potential applied onto the end electrodes $V_{end}$ and the effective potential seen by the ions \cite{champenois09}. $\kappa$ is smaller than 0.02 in our experiment.
For $2k=4$, the two contributions to the radial  effective potential are quadratic and no major impact is expected from the deconfining effect of the axial confinement on the radial trapping. On the contrary, for $2k=8, 12 \ldots$, the addition of a $-r^2$-contribution to the rf-pseudopotential  modifies its shape. The axis of the trap becomes an unstable position and the potential minimum is shifted to $r=r_{min}$ defined by
\begin{equation}
r_{min}^{2k-4}=\frac{r_0^{2k-2}}{z_0^2}\frac{16 {\mathcal E}_k \kappa V_{end}}{(k-1)q V_0^2}. \label{eq_rmin}
\end{equation}
Taking into account this shift of potential minimum is relevant when few cold ions are trapped, as they fill the bottom of the potential well. When large samples are considered, and following Dubin and O'Neil concerning the thermal equilibrium states of trapped non-neutral plasmas \cite{dubin99}, one can show that the density distributions  $n(r)$ are completely controlled by the rf induced effective potential with no contribution from the DC-voltage \cite{champenois09}:
\begin{equation}
\lim_{T\to 0} n(r)=\frac{ \epsilon_0 (k-1)^2 V_0^2}{8 {\mathcal E}_k r_0^2}\left(\frac{r}{r_0}\right)^{2k-4}
\label{eq_densite}
\end{equation}

This distribution  is uniform only for the quadrupole geometry. As mentioned above, for higher order geometries, the density is expected to increase with the distance from the center of the trap, leading to a tube-like self-organized structure. This prediction in the mean-field approach is  confirmed by molecular dynamics simulations for ion numbers above a few hundred \cite{calvo09}.

A detailed discussion of the dynamics and thermodynamics of an ion cloud in a multipole trap can be found in \cite{champenois09}.

\section{Numerical simulations with few ions}
\label{sec:numsim}
The equations of motion of an ion in a multipole trap of order higher than $2k>4$ are non-linear and coupled in the two radial directions $x$ and $y$. Therefore, there are no known analytical solutions and the stability of a trajectory depends on initial conditions as well as trapping parameters. Numerical simulations are then  useful tools to get insight into ion dynamics and to identify the relevant trapping parameters for each  set-up.

We have developed a simulation code which follows the actual trajectories of the trapped ions during a large number of rf periods \cite{marciante10}. This code takes into account all forces that act on the trapped particles including laser-cooling. The simulation can be run with the time-averaged effective potential approach or making use of the local rf electric field. The comparison of these two approaches allows  us to study the influence of rf driven motion on the ion dynamics. Following exactly a large number of ions in the long run   is of course limited by computer performances. At present, all simulations are carried out with less than 1000~ions, higher numbers can be achieved by parallelizing the simulation code which is work in progress.

\begin{figure}[b!]
  \includegraphics[height=0.12\textheight]{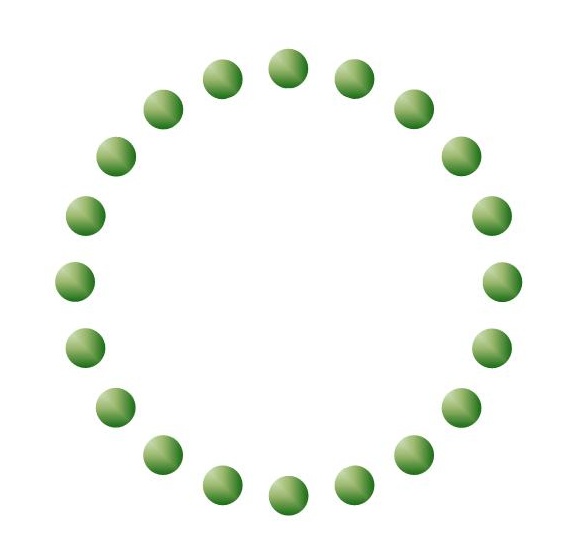}
\caption{Schematic drawing of laser-cooled ions arranged in a ring structure in a multipole radiofrequency trap.}\label{fig:1ring}
\end{figure}

The described code has been used to lead investigations \textsl{a priori} to the experiments. We have shown that for a small number of trapped ions (roughly below 100) and a steep but realistic trapping potential, the trapped particles settle in the potential well in $r_{min}$ to form ring structures (see Eq.\ref{eq_rmin} and Fig.\ref{fig:1ring})  and the number of rings can be controlled by the potential in $z$-direction \cite{marciante12}.  We first considered this ring structure  for frequency metrology applications \cite{champenois10} but recent work suggests that it can also be the realization of a space-time crystal \cite{li12}.

In comparison with the string configuration formed by few ions in a linear quadrupole trap, located along the zero node line of the trapping field, this ring structure suffers from  rf driven motion at the ion equilibrium point. This motion can not be laser cooled and can induce ion-ion collision heating (rf heating).  The advantage of keeping the ions in a single plane is to decouple the axial degree of motion from the radial one \cite{marciante10,champenois10}. The Doppler cooling limit can then be reached along the axial direction despite the radial  rf driven motion. Actually, in contrast to an ion string in a linear quadrupole trap, all ions in a ring will experience identical  conditions. Rings of $N$ ions and a radius of about 400 $\mu$m  could allow the realization of an atomic frequency standard reaching a frequency uncertainty which is comparable to its single-ion counterpart \cite{champenois10}, but with a $\sqrt{N}$ times higher signal-to-noise ratio. This would  assure higher short-term frequency stability and can eventually relax constraints on the stabilization of the clock laser due to shorter probing durations.

\begin{figure}[h!]
  \includegraphics[height=0.2\textheight]{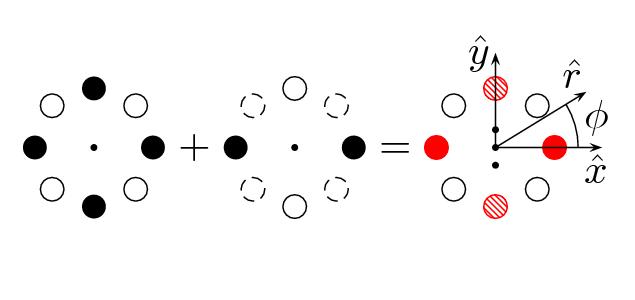}
\caption{Superimposition of a weak quadrupole potential onto an octupole potential. The circles represent the rods of an octupole trap, different colors represent different polarity. The field-free lines are marked by small black dots.}\label{fig:trio}
\end{figure}

Another possibility to avoid the rf-driven motion is the combination of rf electric fields of different geometries in a single device, creating extra field-free regions where ions can be trapped without experiencing micromotion \cite{marciante11}. The general concept is the superimposition of a lower-order multipole potential onto the existing $2k$-pole. A small additional DC voltage will balance ion positions.  Figure \ref{fig:trio} illustrates the example of a weak quadrupole potential  added to an octupole trapping field. Two additional lines of field-free region are created in this case.

\section{Experimental set-up}
\subsection{Design}
\begin{figure}
  \includegraphics[height=0.5\textheight]{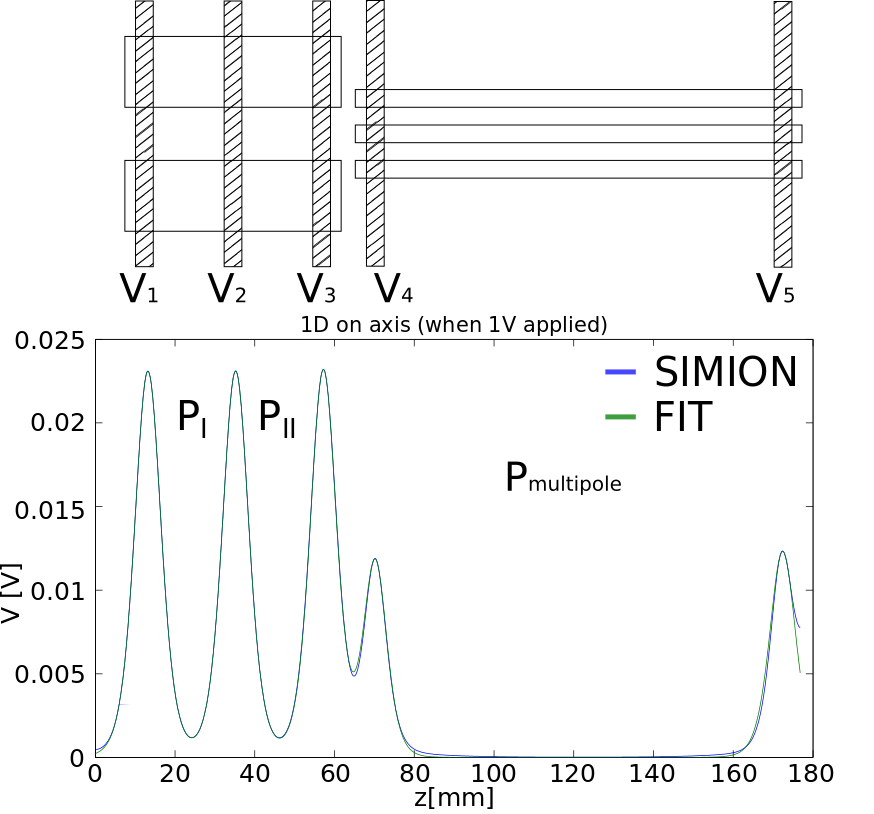}
\caption{DC potential along the $z$-axis of the double ion trap when each DC electrodes is polarized by the same DC voltage.}
\label{fig:trap_1D}
\end{figure}

The simulations  of the small structures described in the previous section demonstrate the underlying processes in the multipole trap, work is underway to extend the numerical approach to large ion numbers, in order to accompany our experiment which is dedicated  to the investigation of the thermodynamics of a large ion cloud in traps of different geometry.  As the dynamics of trapped ions in multipole devices depends on initial conditions, we have been inspired to use a double trap set-up put forward by Prestage and coworkers \cite{prestage07}. Our set-up is, therefore, a combination of a linear quadrupole and a linear octupole trap aligned along a common $z$-axis, the trapping zones being defined by DC voltages applied to extra electrodes. The quadrupole part is itself subdivided in two trapping zones by an additional central DC-electrode, see Fig.~\ref{fig:trap_1D}, in order to have a confinement zone which is exclusively dedicated to ion creation, and a second, clean probing zone, which is not modified by eventual patch potentials. This set-up requires ions to be shuttled from the quadrupole trap to the octupole one.

The device is scaled so that ion numbers larger than 10$^6$ are within reach. To this purpose, we have designed a trap able to store cold ions filling up to a quarter of its size and following as closely as possible the ideal field lines \cite{reuben96,pedregosa10a}. Every defect of the trapping potential will induce higher-order frequency harmonics which are responsible for non-linear resonances that can induce energy transfer from the rf drive to the macro-motion \cite{alheit96}. In addition to geometric considerations, the trapping frequency and amplitude define a working point which keeps rf-heating low. The coupling parameter $\Gamma=q^2/(4\pi \epsilon_0 a k_BT )$ is expressed as the ratio of the average nearest-neighbor Coulomb repulsion energy and the thermal energy. Its usual definition sets $a$  as the Wigner-Seitz radius and is related to the density by $4\pi n a^3/3=1$. In order to be able to observe phase transitions to the Coulomb crystal phase, trap parameters have been dimensioned to reach typical  $\Gamma$ of the order of few hundreds, which adds supplementary constraints on the density of cold samples and requires a minimal stiffness for the radial pseudo-potentials.

As the ratio of the axial to the radial trapping potential controls the aspect ratio of the ion cloud  \cite{hornekaer02},  it is of major importance to assure a sufficiently deep DC axial potential so that the trap can contain dense clouds of the desired size and morphology. This requirement led us to a special design for the DC endcap electrodes which  limits the screening effects of the rf rods \cite{pedregosa10a}. The octupole scaling is matched to the quadrupole one, to make sure that a large cold cloud  fits both trap geometries.

\begin{table}[h!]
\begin{tabular}{|c | c c |}
\hline
  & \textbf{quadrupole} & \textbf{octupole} \\
  &$2k = 4$& $2k = 8$\\
  \hline
trap radius $r_0$	&	3.93 mm		&  4.00 mm			\\
rod radius $r_d$	& 	4.50 mm		&  1.50mm			\\
length $2 z_0$ 		& 2\ * \ 20 mm	& 100 mm			\\
$\Omega/2\pi$ 		&  5.5 MHz		&  3.5 MHz  		\\
$V_{0_{max}}$      & 2500 V$_{pp}$	& 1400 V$_{pp}$	\\
        \hline
\end{tabular}
\caption{Dimensions and trapping parameters of the quadrupole and multipole parts of the trap as described in Fig.\ref{fig:detec_tadoti}.}
\label{tab:param_trap}
\end{table}

Shuttling the ions between the different zones requires  also that both traps share the same rf potential value along their common axis. In practice, this implies that  no pair of rods is grounded; actually,  the rods are polarized  by +$V_0/2 \cos(\Omega t)$ and -$V_0/2 \cos(\Omega t)$, alternately.  Table \ref{tab:param_trap} summarizes the parameters of the double trap device.

\subsection{Ion creation}
Creating large ion clouds requires a fast and non-perturbative method of ionization of calcium atoms. In our device, a medium-size (diameter 4~mm, length  30~mm) oven is electrically heated, the atom beam crosses the trap  almost perpendicularly.
\begin{figure}[h!]
  \includegraphics[height=0.35\textheight]{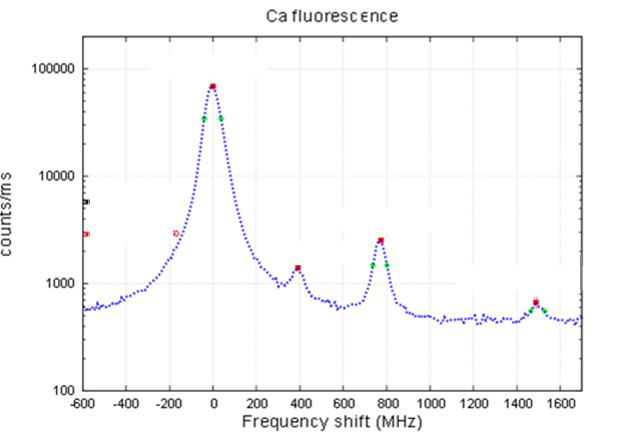}
\caption{Spectrum of the resonant excitation of the atomic calcium beam, showing four different isotopes of calcium. The frequency shift was calibrated from the tabulated isotope shift. }\label{fig:pi}
\end{figure}

 Calcium ions are produced  by two-photon ionization of neutral calcium. Excitation of the resonance line in neutral calcium is the isotope-selective step and requires a tunable laser around 422.7~nm \cite{rotter01,lucas04}. The second step of ionization to the continuum is driven by a simple diode laser emitting at 375~nm. Both laser beams are superposed before injection in the trap along the $z$-axis, which makes their propagation perpendicular to the atomic beam and  the laser-atom interaction is nearly Doppler free.  Indeed, the observed linewidth of the atomic resonance line is only 78~MHz (FWHM), which is less than twice the natural linewidth of this transition (see Fig.\ref{fig:pi}).

\subsection{Ion detection and cooling}
\begin{figure}[h!]
  \includegraphics[height=0.4\textheight]{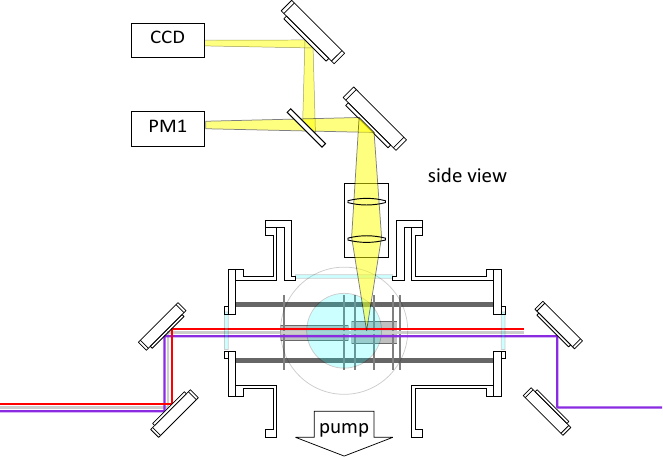}
\caption{Schematic view of the double trap, its detection path, and the laser beams crossing the trap. In this drawing, the octupole trap is situated at the left, and the quadrupole part is the right-hand side of the trap.}\label{fig:detec_tadoti}
\end{figure}

Already during their ionization, ions  are Doppler laser-cooled, using two counter-propagating  collimated  laser beams at 397~nm, aligned along the $z$-axis of the trap. The two laser beams have the same frequency, their intensity is of the order of few mW each, and their waist of the order of 1 mm. Observation of fluorescence from calcium ions requires a repumping laser at 866~nm which is co-propagating with one of the 397-nm laser beams.  A fraction of the photons emitted at 397~nm are collected by a photomultiplier (PM1) and an intensified CCD camera (see Fig.\ref{fig:detec_tadoti}).

\begin{figure}
  \includegraphics[height=0.30\textheight]{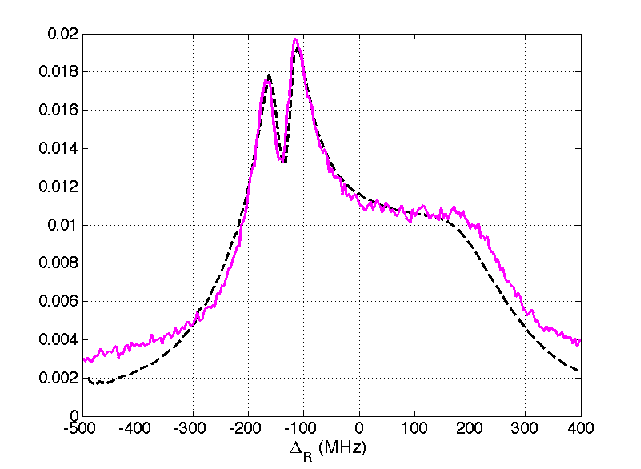}
\caption{Fluorescence spectrum of a warm calcium ion cloud as a function of the  repumping laser (866 nm) frequency $\Delta_R$n with a single co-propagating laser-cooling beam at 397~nm. The magenta line is the normalized experimental data. The black dashed line is the result of numerical simulations for the probability for the ions to be in the excited state. It assumes oscillation amplitudes of the order of 100~m/s and a 397~nm cooling laser detuning of -300 MHz.  The  $\Delta_R=0$ position on the $x$-axis is a result of the simulation.  }\label{fig:ca_fluo} 
\end{figure}

A typical fluorescence spectra from a warm ion cloud is shown in Figure \ref{fig:ca_fluo},  where several distinct features appear. In this case, only one cooling laser is applied, co-propagating with the repumping laser beam. The shape of this Doppler-broadened spectrum can be explained by  the dark resonance generated  from the simultaneous excitation of the 866-nm and 397-nm transitions at Raman resonance, responsible for the double-peak maximum  \cite{lisowski05b}. The large step-like Doppler broadening can be fully reproduced by numerical simulations when the motion of the ions is described by an harmonic oscillation along the laser axis \cite{champenoisxx}.

When laser cooling is efficient enough to compensate the influence of rf heating in the quadrupole trap, a phase transition to a correlated phase (liquid or crystal) \cite{hornekaer02} can be observed by a dip in the fluorescence spectra and/or  by a modification of the shape of the cloud on the camera. When in crystal or liquid phase, the cloud forms a characteristic ellipse whose aspect ratio is controlled by the aspect ratio of the potential and the ion density by the rf pseudo-potential. Comparing pictures of cold clouds in the same rf electric field is then an easy way to compare number of trapped ions as the measurement is independent of the temperature of the ions and the laser detuning.

\subsection{Transport}
One of the experimental challenges of this device is the achievement of a 100\%-transport of a large ion cloud between the different zones of trapping. Moreover, the ions must gain as little energy as possible throughout this transfer.
Similar experiments have been carried out in micro-traps, where single ions in the vibrational ground state are shuttled in order to demonstrate scalable architectures in quantum information processing \cite{kielpinski02}.
Different groups have recently demonstrated diabatic transport of single ions  in micro-fabricated traps, preserving the motional state of the ions \cite{bowler12,walther12}.
Even though objectives and limiting conditions are very similar in our device, the realization is completely different. We aim to transfer a large 3D-ion cloud, over a distance of tens of millimeters, by actuating only three DC-electrodes and eventually the rf potential. The large dimensions of the ion cloud and the trap itself imply the use of much larger rf and DC voltages  than in micro-traps. Similar to micro-traps, DC-voltage switching times are limited by the finite bandwidths of the employed DC drives.

Actually, two conditions are imposed onto the potential during the transport: The axial trapping must not be modified during the transport and the position of the minimum of the potential well, $z_{min}(t)$ follows a given shape, known as the gate function \cite{hucul08}. If both conditions are fulfilled, the energy gain of a single ion due to the transport can be analytically calculated for a given gate function and a given gate duration~\cite{reichle06}. While the second condition is easily achieved, the first one can not be satisfied in our set-up due to the large distances between adjacent DC electrodes. As a consequence the secular frequency of the axial potential experiences a large variation during the transport. To investigate how this variation affects the transport, a numerical code is used, as an analytical approach is not possible. The code uses realistic trap potentials given by SIMION8.1~\cite{simion} to study the transport of ion clouds in our double trap system. It allows  to explore the effects on the transport of many experimental parameters, as a 'gate' deformation due to bandwidth of the power supply, or the effect of time discretization due to the type of Digital-to-Analog-Converter used to generate the waveforms~\cite{pedregosaxx}.

Preliminary experiments of shuttling between both parts of the quadrupole trap are very promising. Transfer efficiencies of 100\% have been achieved for clouds of about 10000 ions. Experiments are ongoing to study the impact of several trapping  parameters before transferring ions to the multipole trap.

\section{Perspectives}
We have recently set up a new trapping device for the study of the dynamics and thermodynamics of large ion clouds. Our aim is to trap clouds composed by more than one million ions. Different features are of interest in this device. Optimization of the experimental parameters includes the investigation of transport properties between different trapping zones. Optimal control protocols might be the solution to shuttle the complete set of trapped ions without energy gain. The investigation of the shape and structure of a large ion cloud in a multipole trap will give insight into the micromotion influence, and the eventual creation of a tube-like structure. This latter constitutes a quasi-2D structure, and may serve as a model system for new applications.


\begin{theacknowledgments}
The authors thank  Michael Drewsen for very stimulating discussions and Vincent Long for his precious help in the conception of the trap and its practical realization. Financial support from ANR (ANR-08-JCJC-0053-01), CNES and region PACA is acknowledged.
\end{theacknowledgments}



\bibliographystyle{aipproc}   


\end{document}